\documentclass{egpubl}
\usepackage{egsr18}

 \ConferencePaper      
 \SpecialIssuePaper         
 \electronicVersion 

\ifpdf \usepackage[pdftex]{graphicx} \pdfcompresslevel=9

\else \usepackage[dvips]{graphicx} \fi

\PrintedOrElectronic

\usepackage{t1enc,dfadobe}

\usepackage{egweblnk}
\usepackage{cite}
\usepackage{gensymb}
\usepackage{float}
\usepackage{soul}
\usepackage{array}
\usepackage{multirow}
\usepackage{hhline}
\usepackage{setspace}
\usepackage{csquotes}
\usepackage{siunitx}

\usepackage[caption=false,font=footnotesize]{subfig}
\usepackage{amsmath}
\usepackage{lineno}
\usepackage{array}

\newcommand*{\new}{}

\newcommand*{\newreb}{}

\DeclareMathOperator*{\argmin}{arg\,min}

\usepackage{datenumber}
\usepackage{calc}

\newcounter{datetoday}
\newcounter{diffyears}
\newcounter{diffmonths}
\newcounter{diffdays}

\newcommand{\difftoday}[3]{%
      \setmydatenumber{datetoday}{\the\year}{\the\month}{\the\day}%
      \setmydatenumber{diffdays}{#1}{#2}{#3}%
      \addtocounter{diffdays}{-\thedatetoday}%
      \ifnum\value{diffdays}>0
        \def\diffbefore{}%
        \def\diffafter{left}%
      \else
        \def\diffbefore{}%
        \def\diffafter{ago}%
        \setcounter{diffdays}{-\value{diffdays}}%
      \fi
      \setcounter{diffyears}{\value{diffdays}/365}%
      \setcounter{diffdays}{\value{diffdays}-365*\value{diffyears}}%
      \setcounter{diffmonths}{\value{diffdays}/30}%
      \setcounter{diffdays}{\value{diffdays}-30*\value{diffmonths}}%
      \diffbefore
      \ifnum\value{diffyears}=0
      \else
        \ifnum\value{diffyears}>1
            \thediffyears\space years,
        \else
            \thediffyears\space year,
        \fi
      \fi
      \ifnum\value{diffmonths}=0
      \else
        \ifnum\value{diffmonths}>1
            \thediffmonths\space months
        \else
            \thediffmonths\space month
        \fi
      \fi
      \ifnum\value{diffdays}=0
      \else
        \ifnum\value{diffdays}>1
            \thediffdays\space days
        \else
            \thediffdays\space day
        \fi
      \fi
      \diffafter
}

\usepackage{soul}

\title[On-the-Fly Power-Aware Rendering] {On-the-Fly Power-Aware Rendering}

\author[Y. Zhang, M. Ortin,  V. Arellano, R. Wang, D. Gutierrez, H. Bao]
{\parbox{\textwidth}{\centering Yunjin Zhang$^{1*}$\ \ \ \ \ \  Marta Ortin$^{2*}$\ \ \ \ \ \  Victor Arellano$^2$\ \ \ \ \ \ Rui Wang$^{1}$\footnote{Corresponding author: rwang@cad.zju.edu.cn}\ \ \ \ \ \  Diego Gutierrez$^2$\ \ \ \ \ \  Hujun Bao$^1$}
        \\
{\parbox{\textwidth}{\centering  $^1$ State Key Lab of CAD\&CG, Zhejiang University\ \ \ \
	$^2$ Universidad de Zaragoza, I3A\\
	*Joint first authors\ \ \ \ $\dagger$Corresponding author: rwang@cad.zju.edu.cn}
}
}

\begin{document}

\teaser{
 \includegraphics[width=\linewidth]{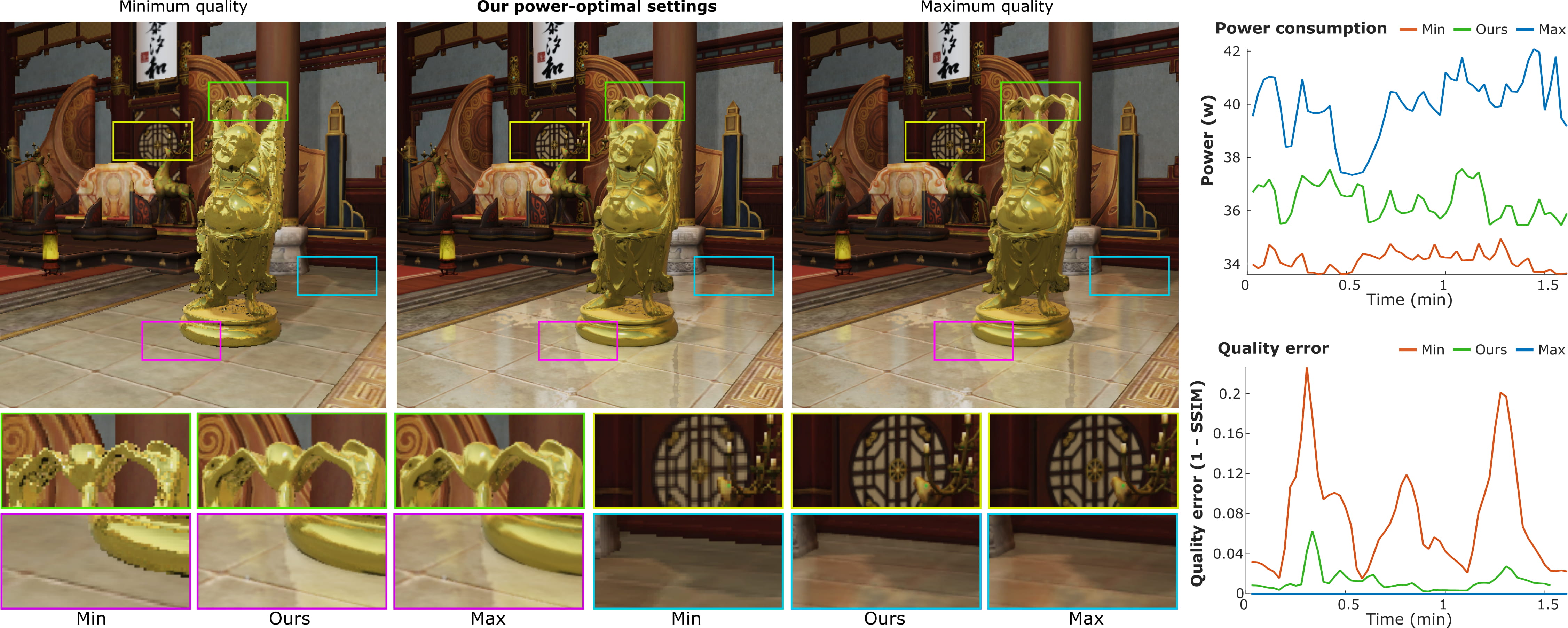}
 \centering
 \caption{We propose a novel on-the-fly, power-aware framework that selects the optimal rendering configuration to maximize visual quality, while keeping GPU power consumption within a power budget. Different from existing approaches, our method requires \newreb{only a few minutes of initialization executed once per platform}. The figure shows results for the \textit{Hall} scene, where our power-optimal settings yield images of similar quality as Maximum Quality, with significantly lower power consumption. The charts on the right show power consumption and quality error (measured with the perceptually-based SSIM metric).}
   \label{fig:teaser}
}

\maketitle
\begin{abstract}
Power saving is a prevailing concern in desktop computers and, especially, in battery-powered devices such as mobile phones. This is generating a growing demand for power-aware graphics applications that can extend battery life, while preserving good quality. In this paper, we address this issue by presenting a real-time power-efficient rendering framework, able to dynamically select the rendering configuration with the best quality within a given power budget. Different from the current state of the art, our method does not require precomputation of the whole camera-view space, nor Pareto curves to explore the vast power-error space; as such, it can also handle dynamic scenes. 
Our algorithm is based on two key components: our novel power prediction model, and our runtime quality error estimation mechanism. These components allow us to search for the optimal rendering configuration at runtime, being transparent to the user. We demonstrate the performance of our framework on two different platforms: a desktop computer, and a mobile device. In both cases, we produce results close to the maximum quality, while achieving significant power savings.

\small{\textbf{Reference:} Yunjin Zhang, Marta Ortin, Victor Arellano, Rui Wang, Diego Gutierrez, and Hujun Bao. On-the-Fly Power-Aware Rendering. Computer Graphics Forum, 2018. doi: 10.1111/cgf.13483}

\ccsdesc[500]{Computing methodologies~Rendering}

\printccsdesc   
\end{abstract}  


\section{Introduction}
\label{sc:introduction}

Current mobile phones and other battery-powered devices incorporate increasingly complex functionalities and applications, which in turn lead to higher power consumptions. Advances in computer graphics have produced highly sophisticated real-time rendering algorithms, which are used in games, data visualization, or virtual reality. 
%
To extend the limited battery life, energy saving becomes a primary goal~\cite{AkenineMoller2008,Johnsson2012}. A lot of research effort has recently been oriented towards characterising the power consumption of rendering algorithms and finding strategies to control the amount of expended energy~\cite{Stavrakis2015,Pool2011,Arnau2014,Wang2016}.


Wang et al.~\shortcite{Wang2016} proposed a state-of-the-art power-saving framework, which traded power consumption for image quality at rendering time. The system was capable of producing high-quality images, while expending significantly less energy. Unfortunately, the system required a pre-processing step of several days, which had to be performed for every different scene to be rendered. Moreover, as a consequence of such precomputation, it could not handle dynamic scenes, and required many memory accesses to fetch the stored data, hampering performance.


In this paper, we propose a  novel real-time, power-saving framework that finds the optimal tradeoff between power consumption and image quality on-the-fly, with only a few minutes of initialization. This is a significantly harder problem, which in turn makes the framework useful for any new game or application without any additional initialization, and enables handling dynamic scenes for the first time. It predicts power consumption and estimates the quality error of different rendering configurations at runtime, and leverages those predictions to adjust the quality level of different shaders, in order to tune the expended energy and keep it within a user-given power budget. 

The main challenge for such real-time power-efficient rendering framework is running 
without affecting user experience. Key to solving this problem are our runtime power prediction and quality error estimation strategies: First, our novel power prediction model (Section~\ref{sc:energy_prediction}) allows us to anticipate the power consumption for every rendering configuration, \textit{without having to measure} the actual energy expended. Second, our quality error estimation (Section~\ref{ssc:error_computation}) obtains the error for all configurations \textit{without the need to render them}. These two components yield extremely accurate predictions, which completely remove the need for the time consuming precomputation of the entire camera-view-space, required for every different scene in Wang et al.'s framework~\shortcite{Wang2016}.

We show results with an in-house, OpenGL prototype implementation that includes six different shaders, with three different quality levels for each one, which yields 729 different shader combinations. We demonstrate the flexibility of our approach by running it on two different platforms: a desktop PC and a mobile device. 



\section{Related Work}
\label{sc:related}

The reduction of power consumption is a growing concern in many different areas, including both algorithms and hardware architecture~\cite{Kyung2014}. Many recent examples have been shown regarding display technology~\cite{Masia2013,Chen2014,Chen2016}, user interfaces~\cite{Dong2009}, or cloud photo enhancement~\cite{Gharbi2015}, to name a few. This issue is specially relevant in mobile devices with limited battery life~\cite{Iyer2003}. We focus here on the particular aspects more closely related to our work: \new{energy saving in rendering and GPU power modeling.}


\textbf{Energy saving in rendering.} 
The power efficiency of several existing graphics algorithms has been extensively examined on different GPUs, as a first step towards reducing the power consumption associated to rendering~\cite{Johnsson2012}. Power limitations are specially relevant in GPUs for mobile devices, and power reduction techniques such as tiling architectures and data compression have been broadly explored~\cite{AkenineMoller2008}. Stravrakis et al.~\shortcite{Stavrakis2015} employ dynamic voltage scaling based on framerate, and implement an energy-aware balancing algorithm that dynamically selects the rendering parameters (geometrical complexity and texture resolution) to save power. Reducing the precision of arithmetic operations can also effectively reduce energy consumption in pixel shaders~\cite{Pool2011}. With respect to hardware-based optimizations, Arnau et al.~\shortcite{Arnau2014} observe that many fragments are repeatedly rendered in different frames, and exploit this redundancy using fragment memoization.

\new{\textbf{GPU power modeling.} 
GPU power can be modeled by considering the static and dynamic power of each one of its architectural units (floating point unit, ALU, cache, memory...) \cite{Hong2010}. Instead, we aim at predicting power consumption using only rendering information, in order to obtain a model directly related to scene complexity.}
Vatjus-Anttila et al.~\shortcite{Vatjus2013}  proposed a model for GPU power consumption taking into account the contributions of three different primitives separately (batches, triangles, and texels), and combining them as a weighted sum. Different from this approach, our model includes render passes, takes into account all primitives simultaneously, includes the number of fragment shader invocations instead of texels as a better predictor of power consumption, and adapts in real-time to changes in the scene. \new{Besides, Vatjus-Anttila et al. need to include an estimated percentage of backfacing and depth culled primitives in order to improve the accuracy of their model. In contrast, we obtain the precise number of primitives used in each stage of the GPU pipeline and use them directly}. This allows us to handle more complex, dynamic scenes, and leads to much higher prediction accuracy.
%
%

Recently, Wang et al. developed a power-optimal rendering framework for mobile devices~\shortcite{Wang2016}, based on Pareto frontiers in power-error space. Despite the very good results, the method requires several days of precomputation of the whole camera-view space for \textit{each} scene. This is impractical, limits the application of the method to static scenes only, imposes large memory requirements, and forces all novel views produced at runtime to be interpolated, which may lead to large errors even in static scenes with large content changes between views (e.g. unoccluded objects).

In contrast, we introduce a novel real-time power prediction model and an error estimation mechanism, which can handle new games and applications without any \newreb{specific} precomputation. It adapts to the scene being rendered in real-time, thus being able to handle dynamic scenes and effects while providing very high accuracy for scenes with different characteristics and complexity.  

\begin{figure}
  \includegraphics[width=\columnwidth]{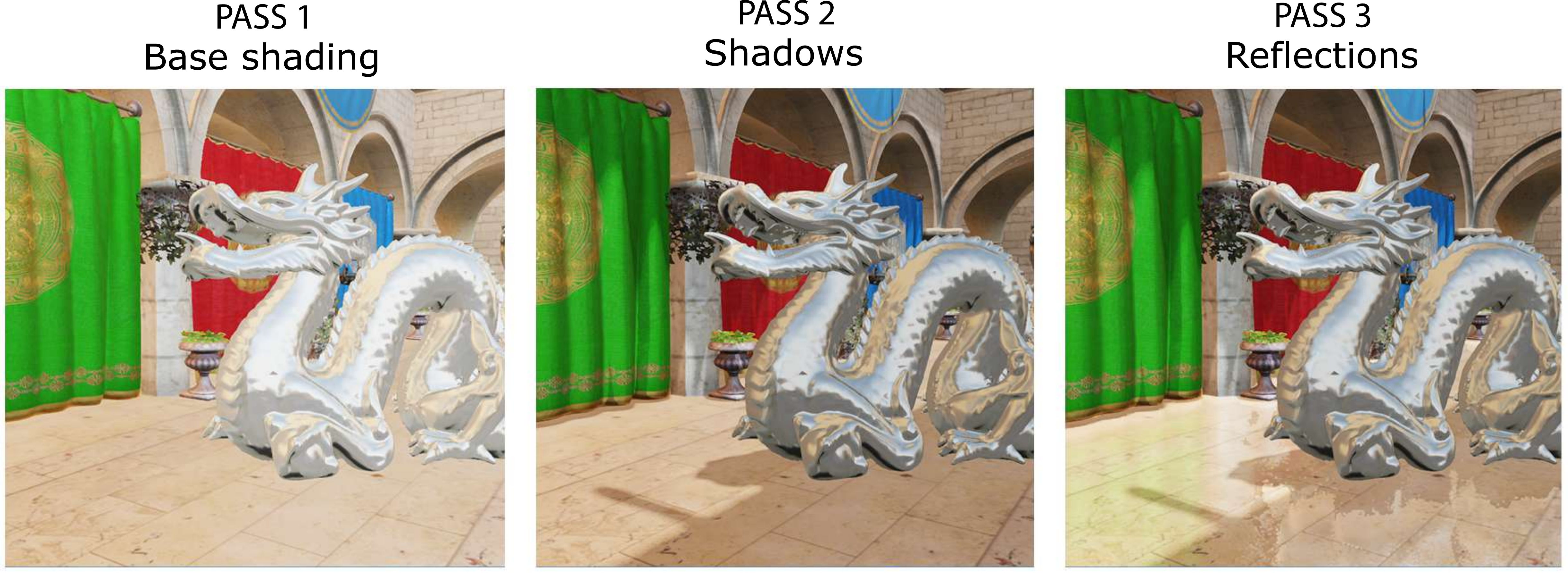}
  \caption{
  Example of the rendering process composed of three rendering passes: base shading, shadows, and reflections. 
  }\label{fig:rendering_passes}
\end{figure}

\section{Problem definition}
\label{sc:problem_definition}
We consider the rendering process as composed of multiple rendering passes that define the visual effects (shadow mapping, reflections, antialiasing...), as  illustrated in Figure~\ref{fig:rendering_passes}. Each pass is executed with a shader with a particular quality level. The input of the rendering process is thus a rendering configuration $\mathbf{s}$ (a vector describing the sequence of shaders corresponding to each rendering pass), and the camera parameters $\mathbf{c}$ (position and view).
In the $\mathbf{s}$ vector, the $i^{th}$ component represents the shader quality level used for the $i^{th}$ pass. 
The contributions from all the passes are combined by a function $f$ to generate the final image; generalizing the rendering process as a function $f$ allows us to implicitly include forward and deferred rendering in our framework. Table~\ref{table:symbols} sums up all the symbols used the paper. 

\begin{table}
\begin{singlespace}
\begin{center}
\begin{tabular}{ | c |p{6.5cm} |}
\hline
  \small $\mathbf{s}$ & \small Rendering configuration: vector with shader quality level for each pass  \\
 \small  $s_i$ & \small Shader for pass $i$ \\
  \small $\mathbf{s_{best}}$ & \small Rendering configuration that generates best quality images \\
   \small $\mathbf{c}$ & \small Camera position and view  \\ 
 \small  $f(\mathbf{s}, \mathbf{c})$ & \small Image rendering function with $\mathbf{s}$ and $\mathbf{c}$.  \\ 
\small   $e(\mathbf{s}, \mathbf{c})$ & \small Image quality error with $\mathbf{s}$ and $\mathbf{c}$, simplified as $e(\mathbf{s})$.  \\ 
  \small $P(\mathbf{s}, \mathbf{c})$ & \small Rendering power with $\mathbf{s}$ and $\mathbf{c}$, simplified as $P(\mathbf{s})$.  \\ 
 \small  $P_{bgt}$ & \small Power budget \\
\small  $P_m$ & \small Minimum power consumption of the GPU \\
  \small $P_M$ & \small Maximum power consumption of the GPU \\
 \small  $b$, $v$, $f$  & \small Batches, vertices, and fragments used to render a frame \\
 \small  $B$, $V$, $F$  & \small  Batches, vertices, and fragments that saturate the GPU \\
 \small  $k_b$, $k_v$, $k_f$  & \small Coefficients for batches, vertices, and fragments \\
  \small $Ins_{vi}$, $Ins_{fi}$ & \small Instructions in vertex and fragment shaders for pass $i$  \\
 \small  $Tex_{vi}$, $Tex_{fi}$ & \small Texel accesses in vertex and fragment shaders for pass $i$  \\
 \small  $\chi$ & \small Cost asociated to the execution of one instruction \\
 \small  $\psi$ & \small Cost asociated to the one texel access \\
\small   $l_j$ & \small Quality level $j$ used for a given pass \\ 
 \small  $l_{max}$ & \small Worst quality level for a given pass \\ 
 \small  $\mathbf{s^0}$ & \small Rendering configuration where every pass uses shader quality level 0, same as $s_{best}$ \\
 \small  $\mathbf{s_i^l}$ & \small Configuration where every pass uses shader quality level 0 except for pass $i$, which uses level $l (l > 0)$ \\
 \small  $\mathbf{s_i^{l_{max}}}$ & \small Configuration where every pass uses shader quality level 0 except for pass $i$, which uses the worst quality level \\
 \small $k$ & \small Coefficient that relates the quality error of two shaders for the same pass\\

\hline
\end{tabular}
\end{center}
 \end{singlespace}
 \caption{Symbols used throughout the paper, and their definition.}
\label{table:symbols}
\end{table}

Different rendering configurations yield  results of varying visual quality. Let $\mathbf{s_{best}}$ denote the rendering settings that generate the best quality image. Similar to the recent work of Wang et al.~\shortcite{Wang2016}, we can define the quality error $e(\mathbf{s},\mathbf{c})$  of any image produced by different rendering settings as 
\begin{align}\label{eq:problem}
e(\mathbf{s},\mathbf{c}) = \int \int_{xy} || f(\mathbf{s_{best}},\mathbf{c}) - f(\mathbf{s},\mathbf{c}) ||  dxy
\end{align}
where $x, y$ define the pixel domain of the image, and $|| \cdot ||$ indicates the chosen norm. 

Besides yielding varying quality errors, different $\mathbf{s}$ and $\mathbf{c}$ vectors also result in different power consumption $P(\mathbf{s},\mathbf{c})$. In general, higher quality images require more power, which generates a tradeoff between power and error. Therefore, given a power budget $P_{\text{bgt}}$, we look for a vector $\mathbf{s}$ such that $e(\mathbf{s},\mathbf{c})$ is minimized, while $P(\mathbf{s},\mathbf{c})$ remains within the budget: 
\begin{align}\label{eq:problem}
\mathbf{s} & = \argmin_\mathbf{s} e(\mathbf{s},\mathbf{c}) && \text{subject to} && P(\mathbf{s},\mathbf{c}) < P_{\text{bgt}}
\end{align}
%

Different from Wang's work, we demonstrate in this paper how to predict $P(\mathbf{s},\mathbf{c})$ and estimate $e(\mathbf{s},\mathbf{c})$ \textit{in real-time}. This is a significantly more difficult problem, since Wang's framework relied on a time-consuming precomputation  (in the order of a few days) of the entire camera-view-space, to be performed for each particular game or scenario.
Our implementation includes six different shaders (resolution, base shading, reflections, shadows, metals, and antialiasing), with three quality levels each, generating a total of 729 different rendering configurations.



\section{Algorithm Overview}
\label{sc:algorithm}

\begin{figure}
\centering
 \includegraphics[width=0.7\columnwidth]{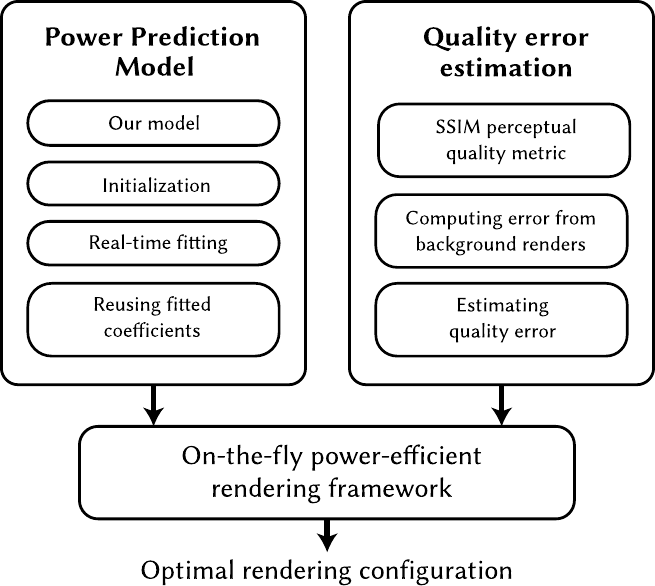}
 \caption{Main components of our algorithm: power prediction model, and quality error estimation. Both components are combined in our on-the-fly, power-efficient rendering framework to generate the optimal rendering configuration.
}
\label{fig:power_error_overview}
\end{figure}

Our algorithm is based on two key components, depicted on Figure~\ref{fig:power_error_overview}: a power prediction model, and a quality error estimation mechanism. Our power prediction model fits a set of coefficients in an equation describing scene complexity, requires minimal initialization, and adapts in real-time to the content being displayed. In order to select the optimal rendering configuration within a power budget, we introduce a strategy to reuse fitted coefficients to predict power consumption in new configurations with minimal computation. 

For our quality error estimation, we first compute the error of a frame with several rendering configurations (one per pass, six in total in our prototype implementation) by running the renders in the background and calculating the SSIM perceptual quality metric. We then use the obtained values to estimate the error for all the other configurations (729 in our case).

Our on-the-fly power-efficient rendering framework makes use of these two components to produce a final image with the highest possible quality, within a given power budget. Sections ~\ref{sc:energy_prediction} and \ref{ssc:error_computation} explain the details of our power prediction model and error estimation mechanism, respectively, while Section~\ref{ssc:shader_selection} describes how the power prediction and error estimation steps are combined at runtime to obtain the optimal rendering configuration.

\section{Power Prediction Model}
\label{sc:energy_prediction}

We introduce our power prediction model based on scene complexity, describe the initialization and real-time fitting process, show how we can predict the power for all our rendering configurations by fitting our model for only one of them, and describe implementation details.

\subsection{Our model}
\new{The rendering pipeline starts running when we command the GPU to draw a group of triangles (called \emph{a batch}) already uploaded to the GPU memory. Those triangles go through several consecutive stages: first, the vertex shader executes per-vertex processing operations; then, rasterization runs per-primitive processing to cull hidden primitives; finally, the fragment shader interpolates per-fragment parameters, texturing, and coloring to generate the final pixel color \cite{BookGPU}.} When setting a fixed frame rate, the complexity of the scene determines the load imposed on the GPU, and power savings are achieved when the GPU is idle between consecutive frames. 
Our power prediction model takes into account consumption at the different stages of the GPU pipeline, according to scene complexity; it includes the number of batches $b$, vertex shader calls $v$, and fragment shader calls $f$ (in the rest of the paper we will refer to these variables as \textit{primitives}).
%

Similar to previous work~\cite{Vatjus2013}, we observe that GPU power consumption follows an inverted exponential function between a minimum $P_{m}$ and a maximum $P_{M}$ power, as the rendering load increases. Given a  rendering configuration $\mathbf{s}$ and camera parameters $\mathbf{c}$, we thus propose the following power consumption model:
\begin{align}\label{eq:power_prediction_model}
P(\mathbf{s},\mathbf{c}) =  P_{m} + (P_{M} - P_{m})  (1 - exp^{-\alpha})  \\
\nonumber  \alpha =  k_b\frac{b}{ B} + k_v  \frac{v}{V} + k_f \frac{f}{F}
\end{align}
 Each one of the $b$, $v$, and $f$ primitives is normalized by the number of elements that causes the GPU to saturate to its maximum capacity ($B$, $V$, and $F$, respectively). They are additionally weighted by coefficients $k_b$, $k_v$, and $k_f$, to take into account the relative impact of each one on the total power consumption. All the parameters depend on the rendering configuration $\mathbf{s}$; additionally, the camera parameters $\mathbf{c}$ are implicitly included in the primitives $b$, $v$, and $f$; we omit these explicit dependencies in the rest of the paper for the sake of clarity.
%

\begin{figure}
    \includegraphics[width=\columnwidth]{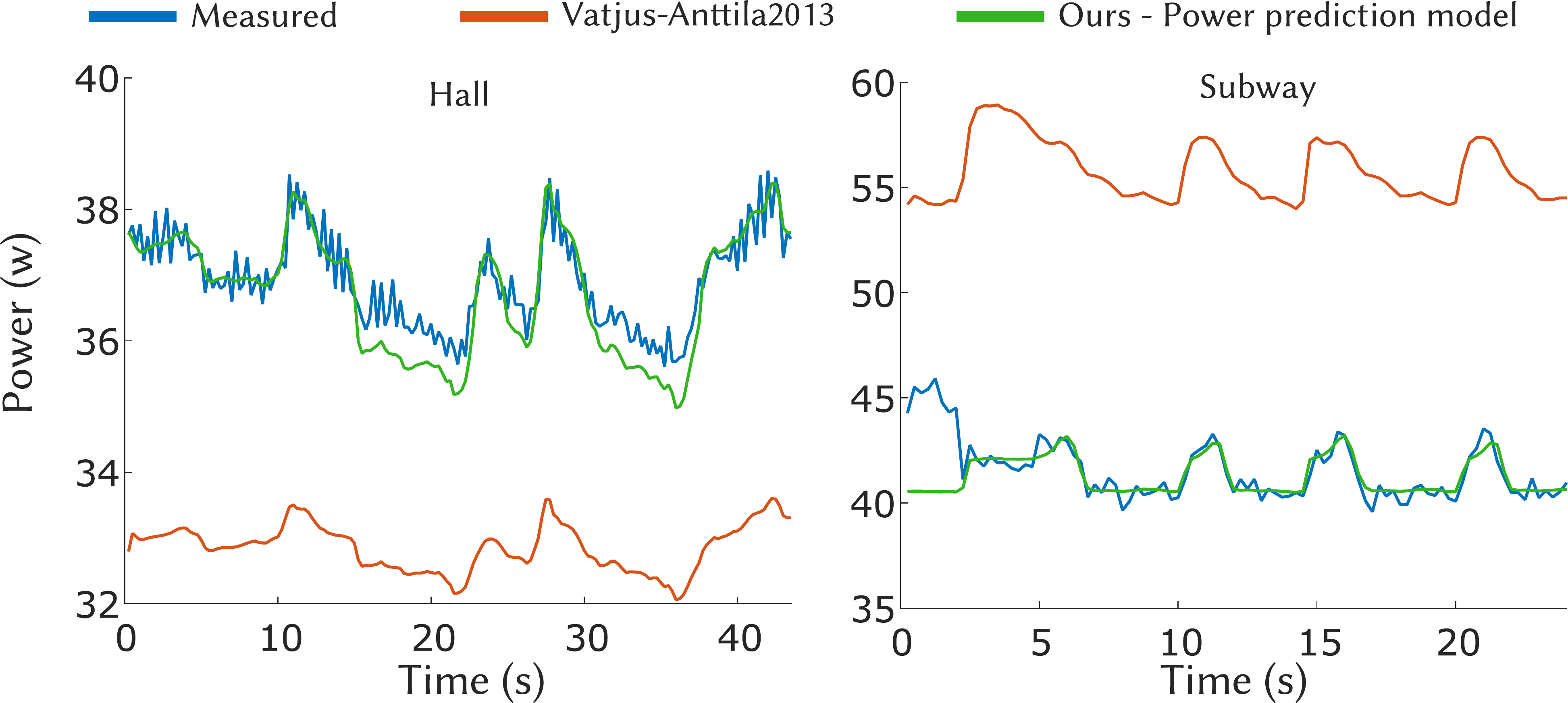}\label{fig:power_generic} \\
    \caption{\new{Power consumption for the \textit{Hall} and \textit{Subway} scenes. Our power prediction closely matches the ground truth, measured power consumption. For comparison, we also show the prediction using the model proposed by Vatjus-Anttila et al. \shortcite{Vatjus2013}.}}
\label{fig:power_model1}
\end{figure}

Since the complete rendering process is composed of several passes, we extend our previous power model to represent each pass individually:
\begin{align}\label{eq:power_prediction_model_extended}
P =  P_{m} + (P_{M} - P_{m}) (1 - exp^{-\sum_i^N\alpha_i}) \\
\nonumber  \alpha_i = k_{bi}\frac{b_i}{B_i} + k_{vi}  \frac{v_i}{V_i} + k_{fi}  \frac{f_i}{F_i}
\end{align}
where $N$ is the number of passes, and the subindex $i$ for each variable indicates its per-pass value. Figure~\ref{fig:power_model1} depicts results for two example scenes (\textit{Hall} and \textit{Subway}) with ground truth measured power, showing how our equation yields a good prediction of power consumption: \newreb{an average of only 1\% error in \textit{Hall} and 2\% in \textit{Subway}}. \new{The model proposed by Vatjus-Anttila et al. provides worse power predictions \newreb{(11\% error in \textit{Hall} and 37\% error in \textit{Subway})} because it does not model the contribution of each rendering pass, it does not consider the number of fragment shader invocations, it uses an estimated correction factor to compute the number of primitives, and it does not adapt to the scene being rendered in real-time. More examples can be found in the supplementary material.}


\subsection{Initialization}
\label{sc:initialization}

Given our power model, we first obtain $P_{m}$ by sending an empty scene to the GPU; we then progressively increase the complexity of the scene to find the values of $P_{M}$,  $B_i$, $V_i$, and $F_i$ that saturate the GPU. This is an offline process that needs to be performed only once per hardware platform, and requires only 3 minutes, in contrast with the costly precomputation of the camera-view-space required for every scene in Wang et al.'s proposal~\shortcite{Wang2016}. In addition, we also obtain the number of instructions and texel accesses for the rendering passes with each quality level, which will be explained in Section~\ref{sc:reusing}.

\subsection{Real-Time fitting}
\label{sc:refitting}
A key aspect of our method is our \textit{real-time power fitting} process, which allows us to obtain very high prediction accuracy without imposing a penalty in performance. When the scene starts running, we collect rendering samples\footnote{A sample includes the measured power for a frame, and its corresponding number of batches, vertices, and fragments for each rendering pass.} during a \textit{fitting window}, and use them to fit coefficients $k_{bi}$, $k_{vi}$, and $k_{fi}$ using a linear regression on the power consumption and number of primitives (see Equation~\ref{eq:power_prediction_model_extended}). 

After the fitting takes place, we check periodically during runtime if the process has to be triggered again, to improve the accuracy of the prediction due to scene changes: First, we collect rendering samples during a \textit{prediction accuracy check window}, and compare our predicted power consumption with the actual consumption measured in the GPU. If the average difference (computed during the frames of the prediction accuracy check window) is above a set threshold, we trigger the real-time fitting process again, and update the coefficients of our power model. This process is illustrated in Figure~\ref{fig:refitting_timeline}.

\begin{figure}
    \includegraphics[width=\columnwidth]{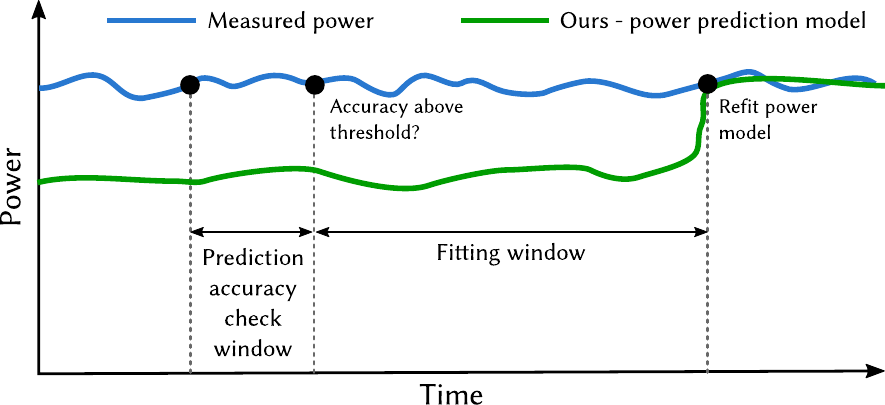}\label{fig:refitting_timeline} \\
    \caption{Timeline showing how refitting works in our model.  First, rendering samples are collected during a \textit{prediction accuracy check} window. We then check if the accuracy of the predicted power is above a given threshold. If the predicted power needs to be updated, we collect new samples during a \textit{fitting} window, which are used to refit the power model and yield a new prediction.}
\label{fig:refitting_timeline}
\end{figure}

\subsection{Reusing the fitted coefficients for other configurations}
\label{sc:reusing}

\begin{figure}
    \includegraphics[width=\columnwidth]{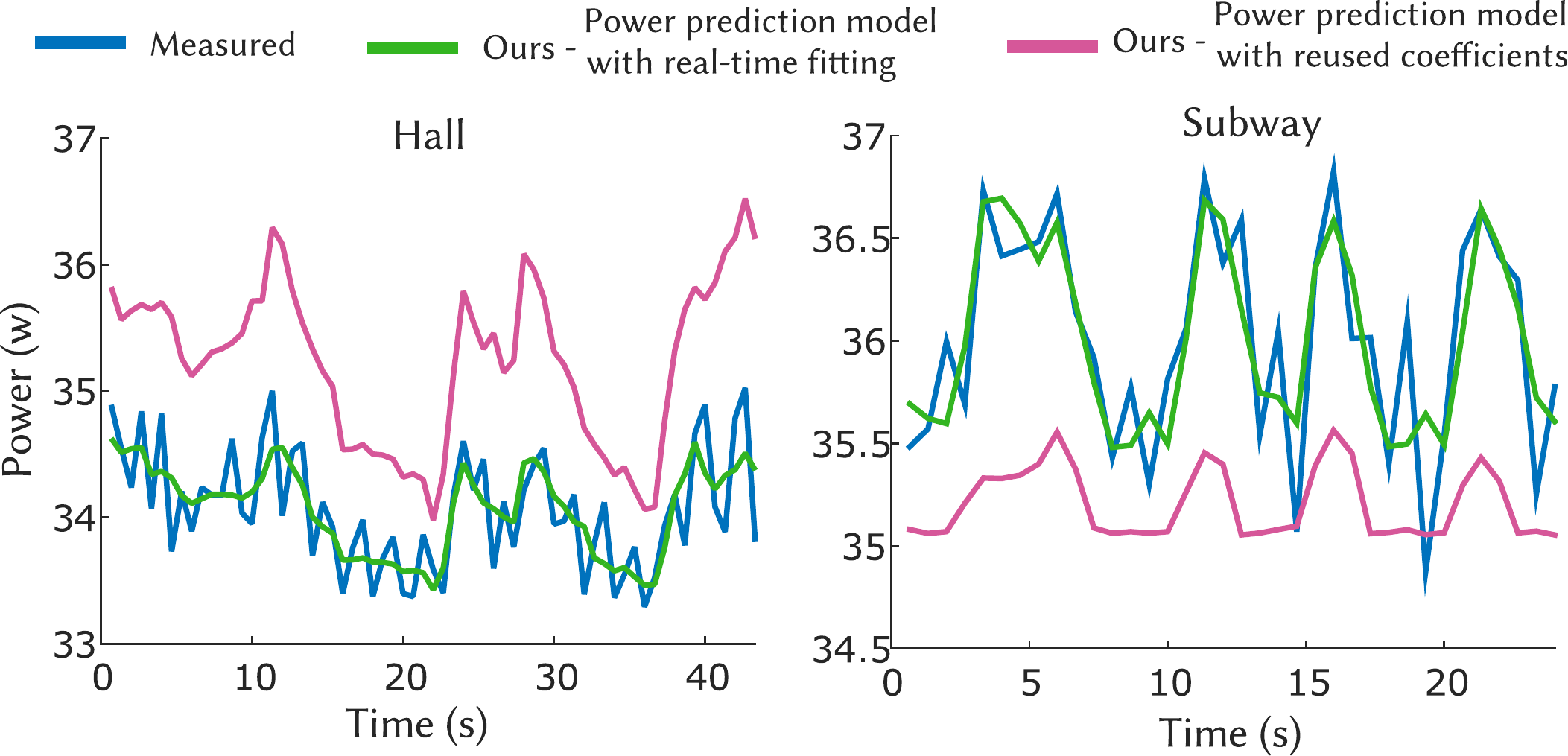}
    \caption{\new{Power consumption of the \textit{Hall} and \textit{Subway} scenes, for rendering configuration $\mathbf{s_A}=(l_0, l_2, l_1, l_1, l_2, l_2)$. We show  ground truth measured data, predicted power with our real-time fitting, and predicted power with coefficients reused from the fitting of a different configuration $\mathbf{s_B}=(l_1, l_1, l_1, l_1, l_1, l_1)$.}}
\label{fig:power_model_refitting}
\end{figure}

At any given moment, we are rendering the scene and fitting our power model with a single rendering configuration. \new{However, each configuration leads to a different power consumption. Therefore, in order to find the optimal rendering configuration, we need to predict the power consumption for \textit{every one} of them (729 in our prototype implementation); fitting the model for every configuration would obviously be impractical, and too computationally expensive.}

To solve this problem, we leverage what our coefficients represent: $k_{bi}$, $k_{vi}$, and $k_{fi}$ express the cost associated to batches, vertices, and fragments, respectively. In particular, $k_{bi}$ is the cost of a rendering request and all the exchange of information between CPU and GPU required to perform the rendering task. In general, this cost is fixed for all the shader quality levels of a given pass, so we can reuse the same $k_{bi}$ for all our rendering configurations. On the other hand, $k_{vi}$, and $k_{fi}$ are related to the number of executed instructions and texel accesses \footnote{\new{Even though any memory fetch could be issued from vertex and fragment shaders, we use the term \emph{texel access} because they generally constitute the vast majory of memory fetches.}}, so we can express the coefficients associated to pass $i$ with shader quality level $s_i$ as:

\begin{align}
\label{eq:vertices_cost}  k_{vi}(s_i) = \chi Ins_{vi}(s_i) + \psi Tex_{vi}(s_i) \\
\label{eq:fragments_cost} k_{fi}(s_i) = \chi Ins_{fi}(s_i) + \psi Tex_{fi}(s_i)
\end{align}
where $Ins_{vi}(s_i)$, and $Ins_{fi}(s_i)$ represent the average number of executed instructions for vertices and fragments in pass $i$, with shader quality level $s_i$; $Tex_{vi}(s_i)$ and $Tex_{fi}(s_i)$ are the average number of texel accesses for vertices and fragments in pass $i$ with shader quality level $s_i$; $\chi$ and $\psi$ are the costs associated to an instruction and a texel access. Since vertex shaders usually have the same number of instructions \new{per triangle} for each pass regardless of the quality level, and they do not access texels, we can simplify Equation~\ref{eq:vertices_cost} as $k_{vi} = \chi Ins_{vi}$.
\new{We obtain $Ins_{vi}$, $Ins_{fi}(s_i)$, and $Tex_{fi}(s_i)$ during the initialization step, which is performed only once per platform (Section 5.2). We instrument the shaders while they are being loaded into the GPU to include atomic counters that automatically count the number of executed instructions and texel accesses for each primitive, and compute the average after running a dummy scene for a few minutes\footnote{\newreb{Any scene can be used to obtain the necessary information, as long as all the shaders in the rendering engine are executed. At 30 fps, running the scene for a couple of minutes allows us to obtain stable, averaged results.}}.} Therefore, the only unknowns are $\chi$ and $\psi$, which we obtain by solving the inconsistent overdetermined system of equations with a linear regression. 


This strategy has one key advantage: By fitting only one rendering configuration, we obtain the coefficients $k_{bi}$, $\chi$, and $\psi$, which do not depend on that particular rendering configuration. We can then reuse them together with Equations~\ref{eq:power_prediction_model_extended}, \ref{eq:vertices_cost} and \ref{eq:fragments_cost}, to obtain power predictions for all our configurations. Thus, the power consumptions associated to different rendering configurations depend only on the number of executed instructions, texel accesses, and primitives. Figure~\ref{fig:power_model_refitting} shows our resulting prediction reusing coefficients from a different configuration.

\subsection{Implementation details}
\label{sc:imp_refitting}

Our prediction accuracy check window has a length of 10 frames, and our refitting window lasts 30 frames. These values are selected to enable a fast fitting process while collecting enough data to ensure the robustness of our fitted model. The complete fitting process and reuse of coefficients is completed in less than 1.5 s after the prediction accuracy check is launched. We set the accuracy threshold to 10\% of the difference between $P_m$ and $P_M$.
%
The linear regression to fit our power model takes an average of 2.6 ms, and the computations to reuse the coefficients for other configurations require 0.7 ms. To ensure that they do not interfere with the rendering process, we execute them in a separate thread on the CPU, while the GPU continues rendering the scene.

\section{Quality Error Estimation}
\label{ssc:error_computation}


To select the optimal rendering configuration, we need to assess the quality error of any frame, by comparing it with its corresponding reference frame  $\varphi_r$, rendered with the highest quality. Similar to Wang's budget rendering framework~\shortcite{Wang2016}, we use the perceptually-based Structural Similarity Index (SSIM)~\cite{Wang2004}, with error given by $e = 1 - SSIM$. 

Let $\varphi$ be the current frame for which we want to obtain the error, and let $\varphi_s$ represent all other alternative renderings using all other rendering configurations. 
In Wang's previous offline approach, every high-quality reference frame $\varphi_r$, all their alternative renderings $\varphi_s$, and their associated quality errors had been precomputed in advance, based on a dense partitioning of the camera-view space of the scene. We face a much harder problem, since we aim to perform all necessary computations at runtime, on a dynamic scene. 
This involves, apart from obtaining the reference frame $\varphi_r$, rendering frames $\varphi_s$ with all other rendering configurations, and calculating their associated quality error. In the rest of the section, we first describe how error is \textit{computed}; however, given the large space of all rendering configurations, it is impossible to compute the error for all of them without visibly affecting performance. We thus introduce our approach to accurately \textit{estimate} most errors, without the need to explicitly compute them.

\subsection{Computing quality error}

Since error computation should not interfere with the user experience, $\varphi_r$ and all $\varphi_s$ are rendered in the background, to a secondary frame buffer (not shown on the screen); $\varphi_r$ is saved in a texture, while each $\varphi_s$ is rewritten in successive renderings after its associated error has been calculated. To avoid a visible drop in the frame rate from rendering $\varphi_r$ and all $\varphi_s$ consecutively, we distribute the task over time. We save the rendering settings used to obtain $\varphi$, as well as the positions of moving objects\footnote{In our implementation, we identify moving objects by the presence of animated skeletal meshes.}, and restore them with an \emph{error computation frequency} to render one frame in the background. 
%


Distributing the rendering tasks over time avoids a sudden drop in performance, but in turn it makes the process excessively long for all the different rendering configurations. To overcome this, the quality error can be computed for just a small subset of rendering configurations. Since this step takes place after power prediction, such configuration subset can be selected from the configurations with higher power below the threshold, which are more likely to produce high quality images.  The selection of the optimal rendering configuration would then choose the best configuration among the available ones. Alternatively, we propose an approximation to obtain \textit{estimated} quality error values for \textit{all} the configurations, without the need to compute all of them. This approximation is suitable for our application, since it allows us to obtain relative estimations to compare different configurations.


\subsection{Estimating quality error for all rendering configurations}
\label{ssc:qual_est}

%
We make two important observations that allow us to \textit{estimate} the error for all 729 rendering configurations by rendering and computing the error for \textit{only six} of them (one per rendering pass).

For the following discussion, we define $\mathbf{s^0}$ as the rendering configuration where every pass uses the highest shader quality (level 0, $l_0$), and $\mathbf{s_i^l}$ as the configuration where every pass uses $l_0$ except for pass $i$, which uses level $l (l > 0)$. Our two observations are:
\begin{itemize}

\item \new{First, we can approximate the quality error for a rendering configuration by adding up the error introduced by each individual pass.} This means that the total error for any rendering configuration can be expressed as a sum of errors using only $\mathbf{s_i^l}$ rendering configurations. For example, with three rendering passes, the error for rendering configuration $\mathbf{s} = (l_2, l_0, l_1)$ can be obtained as:
\begin{eqnarray}\label{eq:example}
e(\mathbf{s} = (l_2, l_0, l_1)) = e(\mathbf{s_0^2}) + e(\mathbf{s_2^1})
\end{eqnarray}

\item Second, given two rendering configurations, $\mathbf{s_i^{l_1}}$ and $\mathbf{s_i^{l_2}}$,  with best quality shaders except for one pass $i$ using shaders $l_1$ and $l_2$, their associated quality errors follow: 
\begin{eqnarray}\label{eq:example2}
e(\mathbf{s_i^{l_1}}) = k e(\mathbf{s_i^{l_2}})
\end{eqnarray}
The set of all coefficients $k$ depends only on the rendering engine used, not on the particular scene being rendered, and thus can be computed beforehand (together with the initialization of our power model). 
\end{itemize}

\begin{figure}
    \includegraphics[width=\columnwidth]{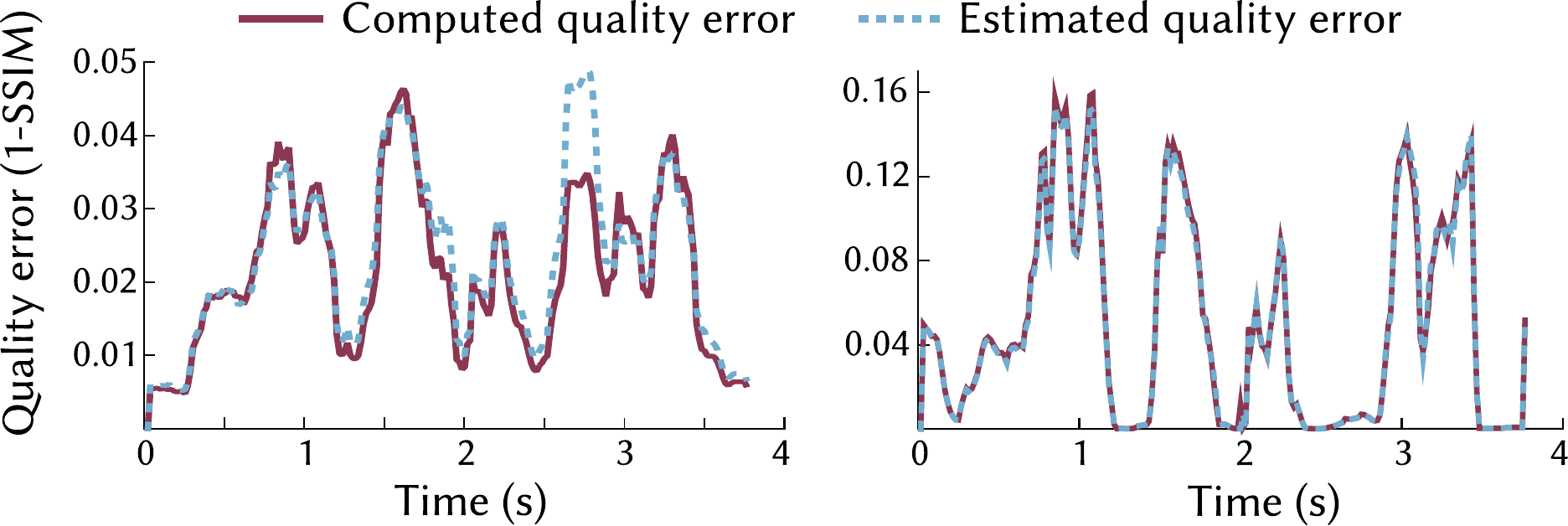}
    \caption{Computed and estimated quality error for the \textit{Subway} scene, using our observations in~\ref{ssc:qual_est}. \textbf{Left:} Quality error for rendering configuration $\mathbf{s} = (l_0, l_2, l_2, l_1, l_0, l_0)$, \new{approximated by adding up the error from each individual pass} (Equation~\ref{eq:example}). \textbf{Right:} Quality error of one rendering pass with medium quality \newreb{(configuration $\mathbf{s} = (l_0, l_0, l_0, l_1, l_0, l_0)$)}, using the observation in  Equation~\ref{eq:example2} and coefficient $k$ obtained from the \textit{Sponza} and \textit{Valley} scenes. Please refer to the digital version to distinguish the overlapping computed and estimated quality errors.}
\label{fig:error_observations}
\end{figure}

Combining these two observations, we can estimate the quality error for all our configurations by computing only the error for all $s_i^{l_{max}}$, that is, one configuration per pass. Figure~\ref{fig:error_observations} shows the accuracy of the estimated error using these simplifications. \new{Note that all error computations and estimations are performed in real-time; only the $k$ coefficients have to be obtained beforehand.}


\subsection{Implementation details}
\label{ssc:imp_error_computation}
We compute the error for a frame $\varphi_s$ once very 10 frames. This error computation frequency was selected to minimize the length of the error computation and estimation process while guaranteeing that the GPU is able to keep up with the target frame rate. Alternatively, the error computation frequency can be adjusted at runtime based on the current and target frame rates. 

Obtaining the SSIM index is computationally expensive, taking an average of 0.05 s. Therefore, after a frame $\varphi_s$ has been rendered in the background, the quality error with SSIM is computed in parallel on a separate thread, while the GPU continues rendering the game. \newreb{The GPU-CPU communication takes 0.02 seconds in the worst case, which corresponds to the exchange of data to compute the SSIM index for 2048x2048 resolution.}

\section{On-the-fly Power-Efficient Rendering }
\label{ssc:shader_selection}

\begin{figure}
  \includegraphics[width=\columnwidth]{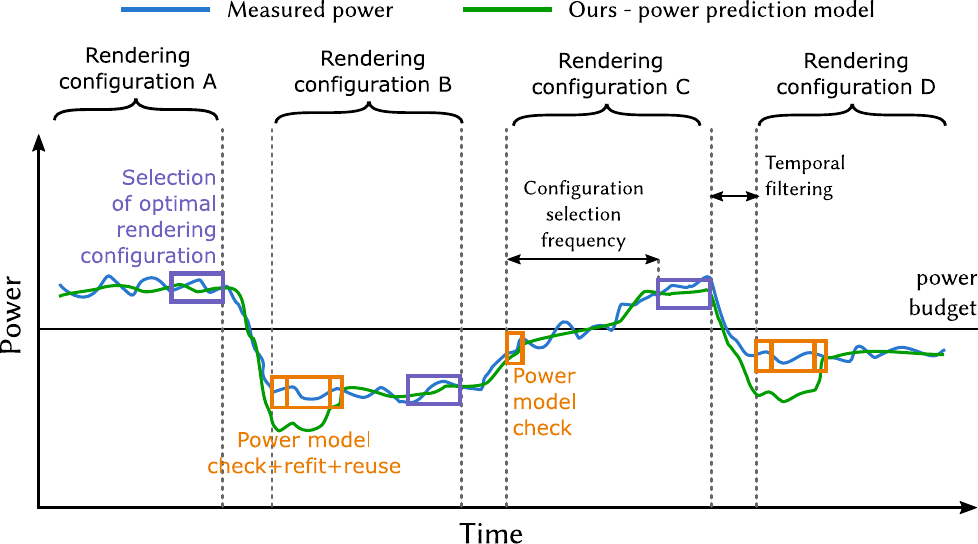}
  \caption{Timeline illustrating power consumption during rendering (measured and predicted with our model), and how our algorithm is executed. Given a configuration selection frequency, a new optimal rendering configuration is selected (purple box). Our temporal filtering is executed to transition to the new configuration. Immediately after that, the power accuracy check is performed, followed if necessary by the real-time fitting, and the reuse of fitted coefficients (orange boxes). In this example, when switching to rendering configuration B, the power check step detects an above-threshold gap between the measured and the predicted power, so fitting and reuse are activated. However, when switching  to rendering configuration C, the power check confirms that the gap is below threshold, and refitting and reuse are not launched. Please refer to the text for more details.}
  \label{fig:overview_timeline}
\end{figure}

In the previous sections we have described our power prediction model and quality error estimation mechanism. We now show how those components are combined at runtime to select the optimal rendering configuration. Our periodic selection for the optimal configuration is followed by a temporal filtering to gradually transition to the new configuration, as illustrated in Figure~\ref{fig:overview_timeline}. When the new configuration is set, we start the real-time fitting of our power model.


\subsection{Selection of the optimal rendering configuration}

Given our power predictions and quality error estimations, we aim to find the optimal rendering configuration for a given scene and camera parameters, minimizing quality error while meeting our power budget, as formulated in Equation~\ref{eq:problem}. 
This selection process is triggered periodically, with a ~\emph{configuration selection frequency}. 

We first use our power prediction model to obtain the power consumption for the current frame with all possible rendering configurations. Wang et al.~\shortcite{Wang2016} precompute the power and error for all configurations, producing a large two-dimensional power-error space. To simplify their runtime search for the optimal configuration, they also precompute the Pareto frontier to reduce their two-dimensional exploration of the power-error space to a one-dimensional search along the Pareto frontier. Instead, we predict the power consumption and estimate the error at runtime. This is difficult, as argued in the paper, but in turn it offers an additional advantage: since the power budget is known in advance, when we predict power consumption we can discard all the configurations with a power consumption higher than our budget. The costly two-dimensional search in power-error space is then reduced to a one-dimensional search in error space; this means that we can completely eliminate the need to compute the Pareto frontier.

\begin{figure}
    \includegraphics[width=0.7\columnwidth]{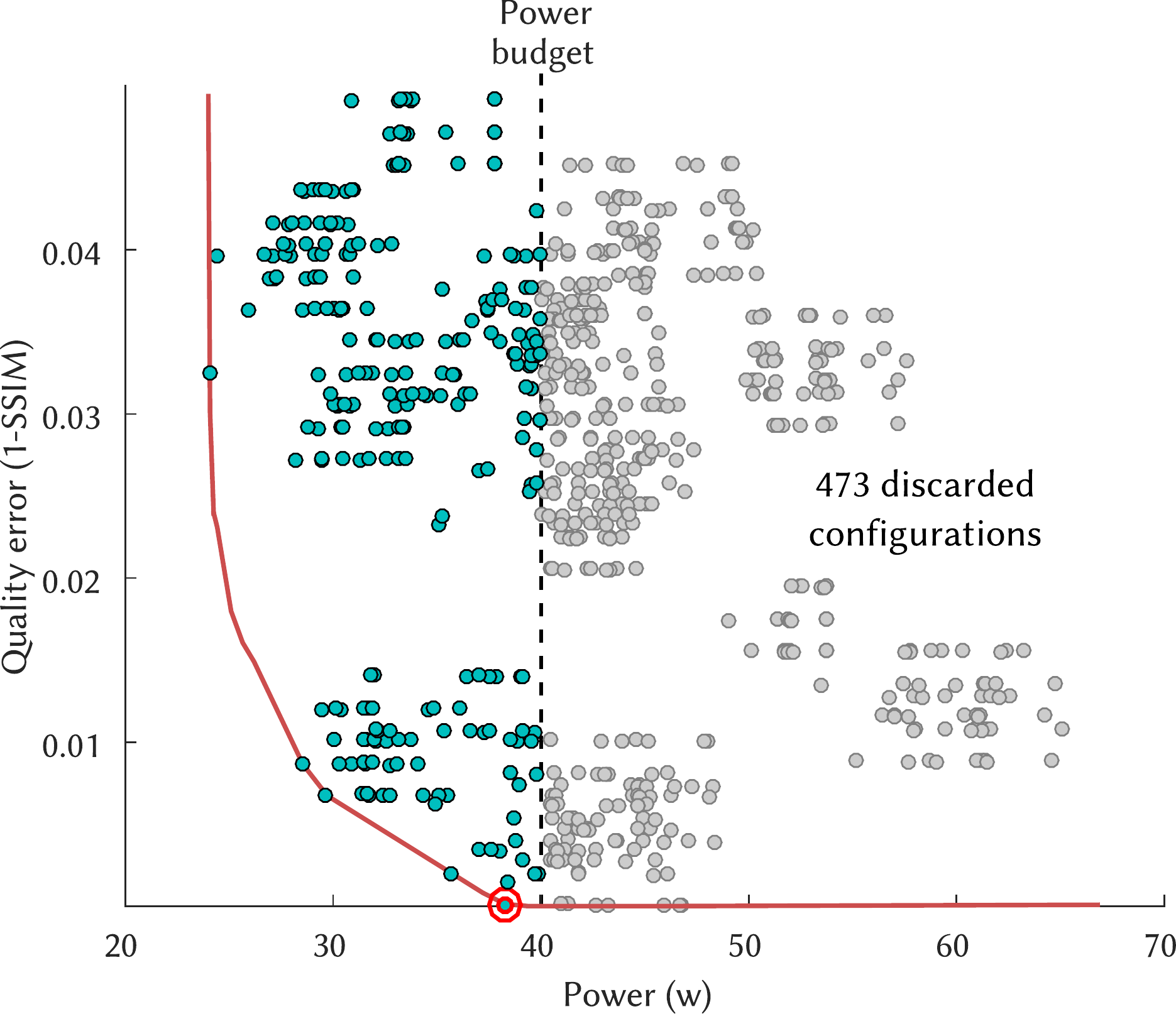}
    \caption{Our rendering configurations drawn in power-error space. We discard all the configurations over the power budget, and perform just a one-dimensional search in error space on the remaining configurations, selecting the one with lowest error (marked with a red circle). This eliminates the need to compute the Pareto frontier in our framework (shown in red for reference only). }
\label{fig:pareto}
\end{figure}

For the configurations that meet our power budget, we estimate the quality error following the process described in Section~\ref{ssc:error_computation}: We render $\varphi$ in the background with configurations $s^0$ and $s_i^{l_{max}}$, and compute their error (according to the error computation frequency to ensure a constant frame rate), and use Equations \ref{eq:example} and \ref{eq:example2} to estimate the quality error for the rest of the configurations. Finally, since we already discarded all the configurations above the power budget, we simply need to choose the rendering configuration with the lowest quality error. This process corresponds to the purple box in Figure~\ref{fig:overview_timeline}, and  is illustrated by Figure~\ref{fig:pareto}, which shows how our rendering configurations are distributed in power-error space, and how our strategy is effective in selecting the one with lowest error within the power budget.

Since we do not rely on scene-specific precomputed data, and different scenes may have very different power requirements, which are thus not known in advance, setting the power budget as an absolute predefined value (as in Wang et al.'s proposal~\shortcite{Wang2016}) is not practical. Therefore, we define the power budget as a percentage between the minimum and maximum power consumption of the scene, which is a very intuitive value to represent the trade-off between power consumption and image quality. For example, if our power budget is 40\%, only configurations with predicted power lower than $P_m + 0.4 (P_M - P_m)$ will be eligible.

\subsection{Temporal filtering}

To avoid a sudden change in image quality when a new rendering configuration is selected, the transition to the new configuration is performed smoothly with the temporal filtering introduced by Wang et al.~\shortcite{Wang2016}. During an interpolation interval T, while the framework transitions from $\mathbf{s_{old}}$ to $\mathbf{s_{new}}$, the effective rendering configuration used for rendering $\mathbf{s_{eff}}$ is computed as:
\begin{eqnarray}\label{eq:temporalfiltering}
\mathbf{s_{eff}} = [(1 - \frac{t}{T})\mathbf{s_{old}} + \frac{t}{T}\mathbf{s_{new}}]
\end{eqnarray}
where the brackets denote the closest integer and $t$ is the time after starting the transition to the new configuration.

\subsection{Real-time fitting of the power model and reusing coefficients}

Every time a new rendering configuration is set, our power prediction accuracy check is triggered (small orange box after temporal filtering in Figure~\ref{fig:overview_timeline}). If the accuracy of our prediction is below a threshold, we refit our power model, as explained in Section~\ref{sc:refitting}. This happens twice in Figure~\ref{fig:overview_timeline}, and is represented by larger orange boxes. The newly fitted coefficients are then used to update the power model for all other configurations by obtaining the cost associated to each instruction and texel access (Section~\ref{sc:reusing} and third small orange box in Figure~\ref{fig:overview_timeline}).

\subsection{Implementation details}
\label{ssc:imp_shader_selection}

The configuration selection frequency triggers the process to select a new optimal rendering configuration 200 frames after the previous configuration was set. This frequency allows us to quickly detect changes in the scene while minimizing the impact of the associated computations. Refitting the power model and reusing the coefficients for other configurations (3.4 ms), predicting the power for all configurations (0.02 ms), and estimating the error and selecting the optimal configuration (0.03 ms) are executed on separate threads. The temporal filtering interval used for interpolation is 2 seconds.


\section{Implementation}
\label{sc:implementation}

To show how our on-the-fly power-budget framework adapts to different hardware, we have implemented it on two different platforms: A desktop PC with an Intel Core i7-7700 and an NVIDIA Quadro P4000, and a mobile Qualcomm Snapdragon 660 (with a 8x Kryo 260 CPU and an Adreno 512 GPU).

%
\subsection{Power Measurement}
To measure the power usage of the graphics card in the desktop PC, we use the NVIDIA Management Library (NVML)~\cite{NVML2015}, which allows us to directly access the power usage of the GPU and its associated circuitry. The specifications report an accuracy of 5\%. In our mobile device, we use an external source meter to directly supply the power of the device. We use a Keithley A2230-30-1, which provides APIs to access the instantaneous voltage and current (same setup used by Wang et al.\shortcite{Wang2016}). During the stages of our algorithm when we have to collect rendering samples (for the power prediction accuracy check and real-time refitting), we measure the power consumption for every frame. In order to reduce variance, in our graphs we report the average power measured over 30 frames.



\subsection{Rendering Configurations}

Our rendering framework runs at 30 frames per second in the desktop PC and at 10 frames per second in the mobile device, and has six passes, each one with shaders of three different quality levels; this amounts to a total of 729 different rendering configurations. The complete set of parameters and values of these shaders is given in Table~\ref{tab:effects}. In particular, we have included: 

\textbf{Resolution:} When setting the resolution of a frame, the number of fragments for other passes are proportionally scaled.\footnote{\new{The resolution is technically not a pass, it sets the screen resolution, which affects other passes. However, it is included in the list of passes for convenience, because it has an effect on power consumption and quality error, and has to be considered as an additional degree of freedom when selecting the optimal rendering configuration.}}

\textbf{Base Shading:} The simplest level is a cheap specular shader, which is improved with a better model for point lights in the next level. The best quality level implements microfacet-based shading.

\textbf{Reflections:} For objects with specular materials; \newreb{it is a multi-pass shader} where quality levels increase the number of generated secondary rays, and the kernel size for color filtering \cite{Stachowiak2015}.


\textbf{Shadows:} The quality level is given by the resolution of the shadow map.

\textbf{Metals:} It is an importance sampling algorithm where quality levels are defined by the number of samples.

\textbf{Antialiasing:} We rely on the FXAA morphological antialiasing to detect edges in the pixel shader \cite{Lottes2009,JIMENEZ2011_SIGGRAPH11}.


\begin{table}[t]
\centering
\small{
\begin{tabular}{|l|c|c|}
  \hline
  \textbf{Passes} & \textbf{Parameters} & \textbf{Values} \\
  \hline
  Resolution & buffer resolution &  60\%, 80\%, 100\%\\
   \hline
  Base shading & specular reflections & cheap spec., improved point \\ &&lights, microfacet spec. \\
   \hline
  Reflections & (samples, kernel) & off, (16,1), (64,9)\\
  \hline
  Shadows & map resolution & 512, 1024, 2048\\
   \hline
   Metals & samples & 2, 6, 60\\
   \hline
  Antialiasing & steps & off, 2, 32\\
  \hline
\end{tabular}
}
\caption{List of parameters and values forming the space of rendering settings.}
\label{tab:effects}
\vspace{-5pt}
\end{table}

When using Equation~\ref{eq:power_prediction_model_extended}, we consider the following: i) The resolution pass has no associated primitives, only having an effect on the number of primitives used in other passes. Therefore, we do not include that pass specifically in the power model formula. And ii) The antialiasing pass works on the final image, and thus does not depend on the number of batches and vertices, it is only affected by the number of fragments.



\section{Results and Evaluation}
\label{sc:evaluation}

We have tested our power-efficient rendering framework on two different platforms (a desktop PC and a mobile device), with four scenes of different complexity, to verify its efficiency in a wide range of  scenarios; refer to Table \ref{table:scenes} for a summary of their main characteristics. In every case, we are able to maintain the predefined 30 frames per second (10 fps in the mobile device).
Our framework supports free exploration of the scene, but we use predefined camera paths to facilitate comparisons and measurements with different qualities, and show the potential of our framework in the long run. For each demo, we specify the preset power budget used to guide our optimal configuration selection process.

\begin{table}
\begin{center}

\small{
\begin{tabular}{|l||c|c|c||c|}
  \hline
  \multirow{2}{*}{Demos} & \multicolumn{3}{|c||}{Scene Statistics} & \multicolumn{1}{|c|}{Rendering}\\
   &  Triang. & Objects & Scene Size & Duration \\  \hline
  Hall & 229.4 k & 23 & 22.33 MB & 1.6 min \\ \hline
  Sponza & 262.1 k & 381 & 25.4 MB & 2.7 min \\ \hline
  Valley & 143.3 k & 61 & 17.0 MB & 1.5 min\\ \hline
  Subway & 526.6 k & 453 & 77.5 MB & 2 min \\ \hline
\end{tabular}
}
\caption{Statistics for our four demo scenes, including number of triangles, number of objects, size on disk of each scene, and duration of the demo.} \label{table:scenes}
\end{center}
\end{table}

\begin{figure}
    \includegraphics[width=\columnwidth]{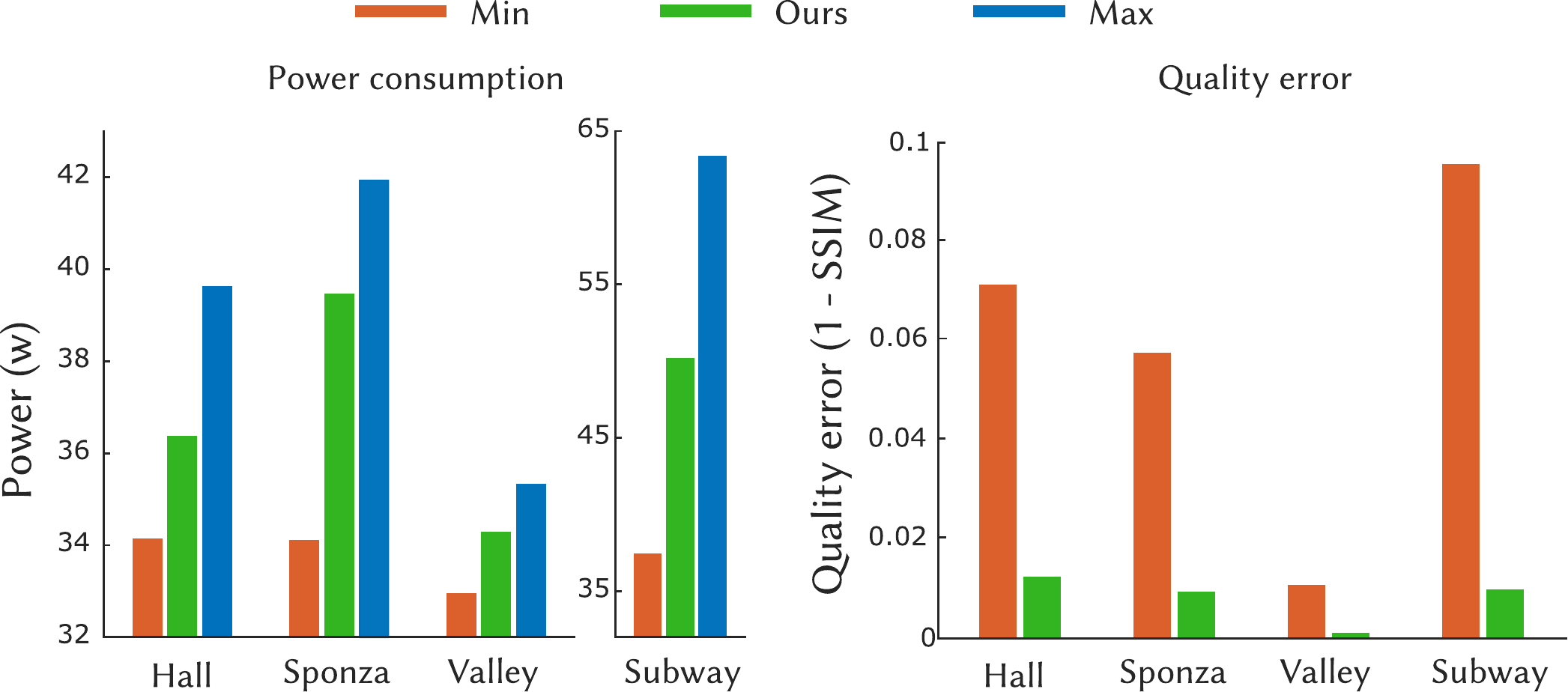}
    \caption{Average power consumption per frame and quality error in our four demos. Note that the quality error for maximum quality is zero.}
\label{fig:power_error_avg}
\end{figure}

Figure~\ref{fig:power_error_avg} shows the average power consumption and average quality error of the four scenes, with maximum and minimum quality, and using our framework with the power budgets reported in this section. It can be seen how we significantly reduce power consumption, while keeping visual quality very close to the maximum.
%
%
%

\begin{figure*}
\subfloat{\includegraphics[width = 0.9\textwidth]{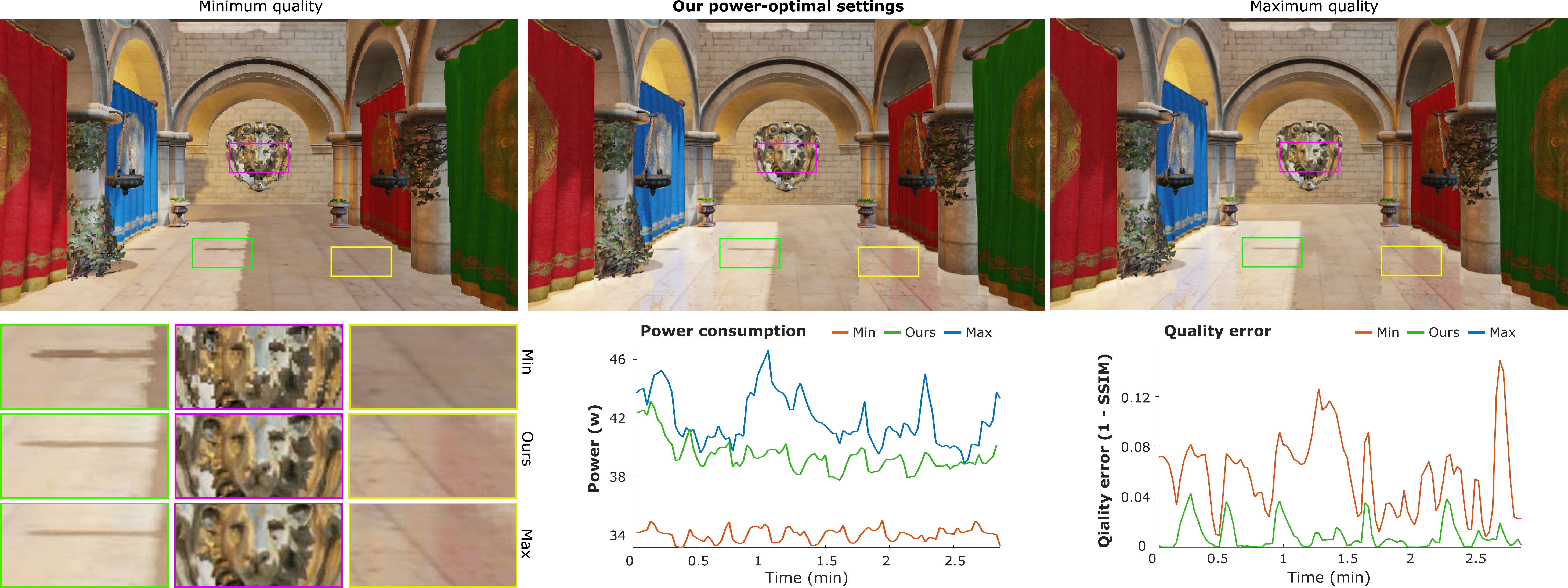}} \label{fig:sponza}
\subfloat{\includegraphics[width = 0.9\textwidth]{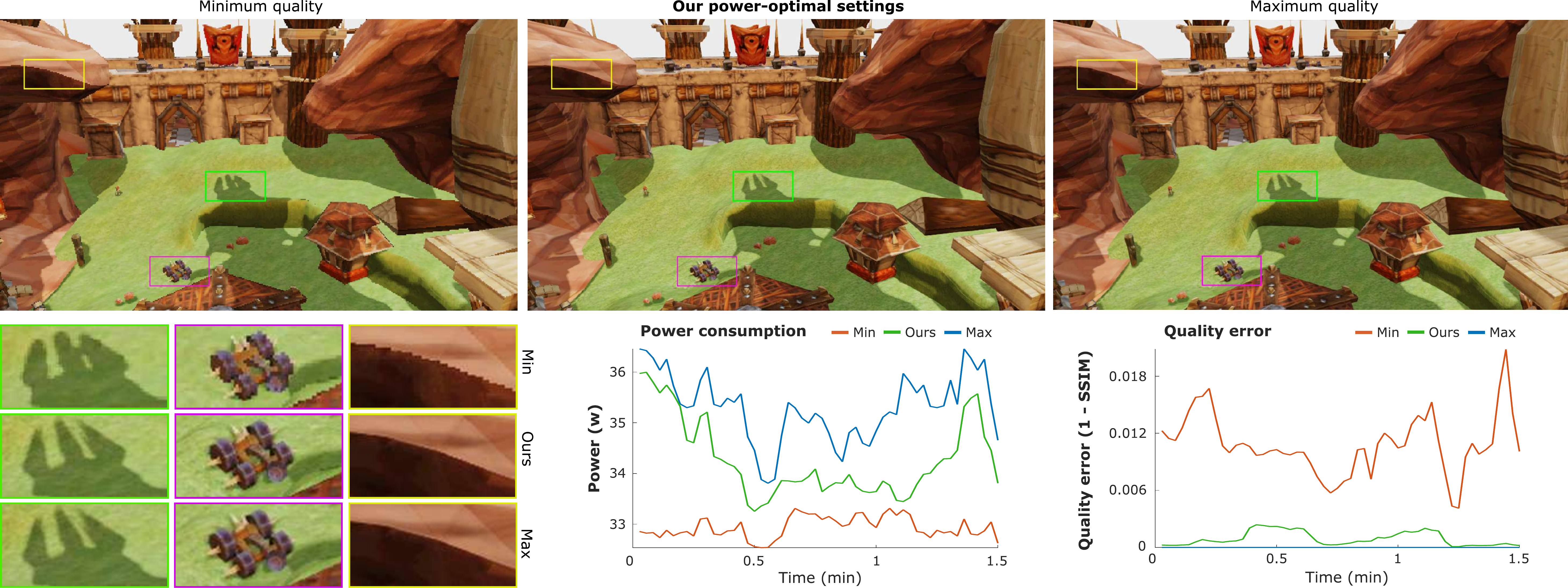}} \label{fig:valley}
\subfloat{\includegraphics[width = 0.9\textwidth]{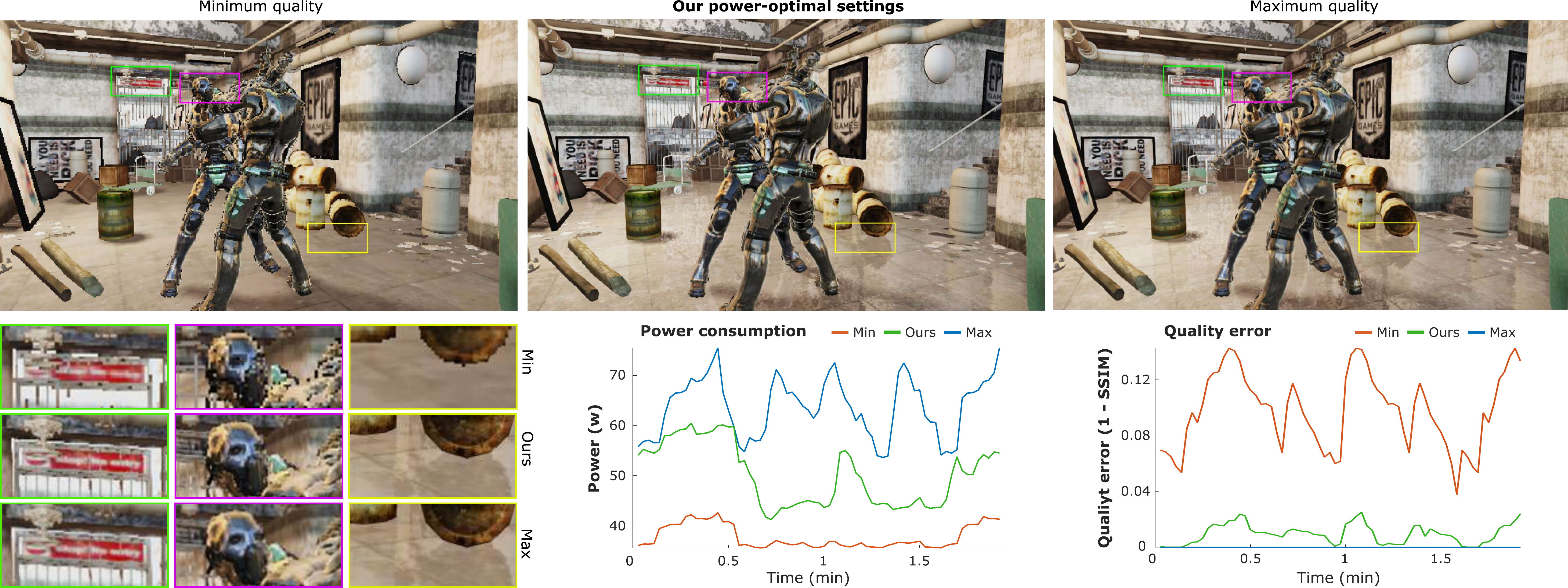}} \label{fig:subway}
\caption{Sponza (top), Valley (middle) and Subway (bottom) demo scenes, executed on a desktop PC. We compare the minimum and maximum quality rendering configurations against our power-optimal configuration. For Sponza, we use a power budget of 60\% \newreb{ (which corresponds to a percentage of the difference between the Min and Max power consumptions)}; for Valley we use 40\%, and for Subway we use 50\%. Our method generates images very similar to those rendered with the maximum quality configuration, while keeping power consumption lower. Please refer to the supplementary video for the full demos.} \label{fig:results_scenes}
\end{figure*}

In the following, we show images from our four scenes with the maximum and minimum quality rendering configurations, together with the result of our power-aware framework. Zoomed-in insets allow to better appreciate details, showing how our results are close to the maximum quality, at a reduced power cost (shown in the accompanying plots). In addition, the supplemental video shows the full demo, including split-screen comparisons.

\textbf{Hall:} This scene has a spotlight acting as a lamp and is composed of  diffuse objects, except for the reflective floor and two metallic buddha statues. It has a high polygon count but very few objects, thus being useful to test scenarios with a small number of batches. Results for our framework running under a power budget of 40\% are shown in Figure~\ref{fig:teaser}. 

\textbf{Sponza:} We use one directional light as the Sun, and one spot light as a candle, both casting shadows rendered by our shadow pass. There are two metallic lion head ornaments (rendered with our metal pass), and the rest of the scene is diffuse (rendered with our base shading pass). The floor of the scene is slightly reflective.
Figure~\ref{fig:results_scenes}, top, shows the results for a power budget of 60\%.

\textbf{Valley:} This relatively low-poly scene is illuminated using the Sun as a directional light, without any spotlights. There are no reflective or metallic objects, hence demonstrating the effectiveness of our model on scenarios where some passes do not affect quality error. This also leads to a smaller difference between maximum and minimum power consumption; although this challenging scene limits the range for improvement, our framework still manages to save considerable energy with minimal image degradation when setting the power budget to 40\%  (Figure~\ref{fig:results_scenes}, middle).

\textbf{Subway:} This complex, high-poly scene is used to test our method in high power usage scenarios. The scene is located underground, so it has no Sun. All the lighting comes from a spotlight located in a lightbulb. The floor of the scene is reflective. The two fighting soldiers are metallic, while the remaining objects are diffuse. The soldiers are animated using skeletal meshes, allowing us to test our framework on a dynamic scene. Results with a power budget of 50\% are shown in Figure~\ref{fig:results_scenes}, bottom.

\newreb{Additionally, in the supplementary material we show the results for different power budgets applied to the \textit{Sponza} scene.}

\begin{figure}
\includegraphics[width = \columnwidth]{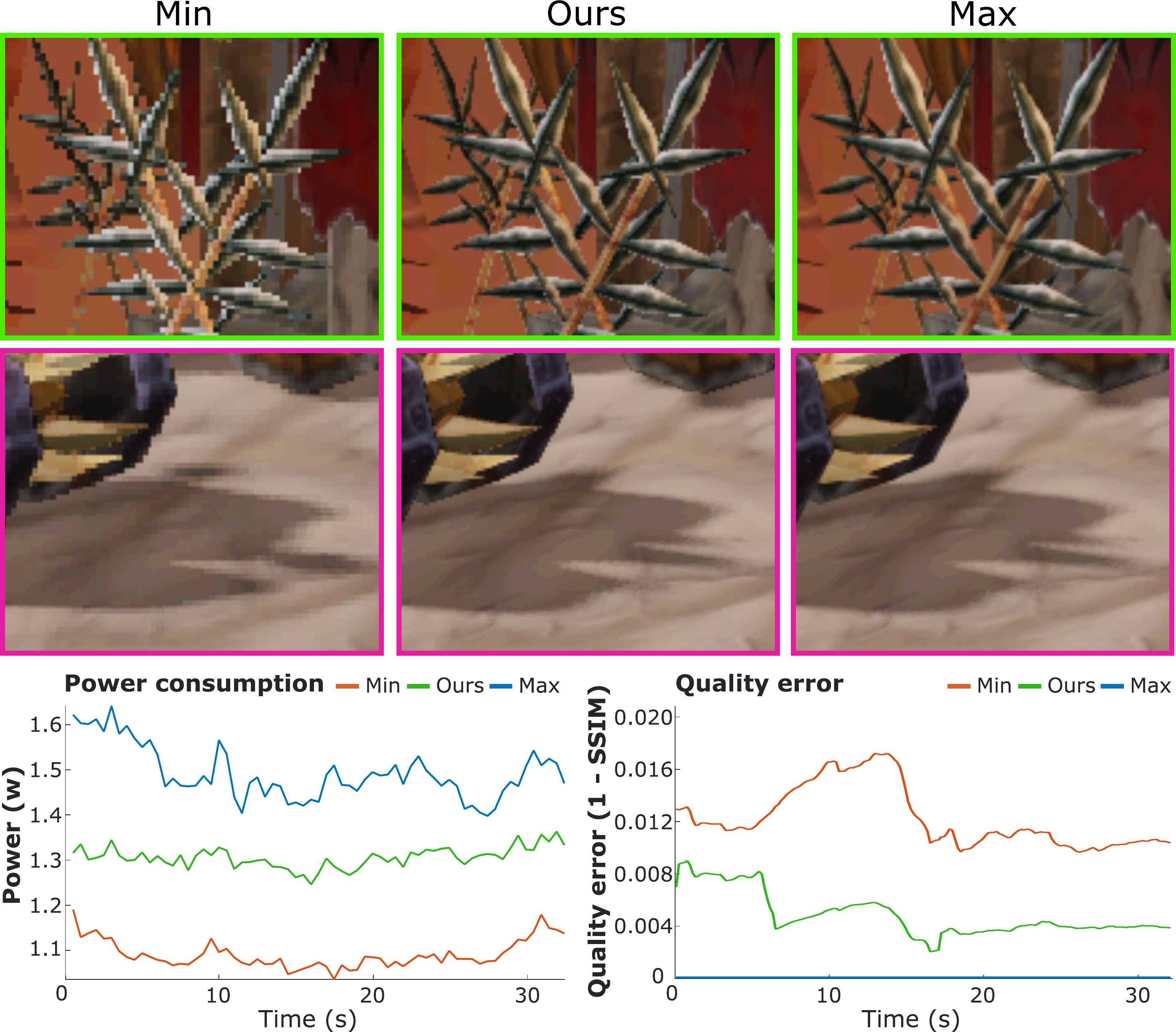}
\caption{Valley demo scene executed on our mobile device. We compare the minimum and maximum quality rendering configurations against our power-optimal configuration with a 50\% power budget. Our method generates images very similar to those rendered with the maximum quality configuration, while keeping power consumption lower. Please refer to the supplementary video for the full demo.} \label{fig:results_scenes_mobile}
\end{figure}

To demonstrate the efficiency of our framework on mobile devices, we show additional results for the \textit{Valley} scene running in our mobile phone, with a power budget of 50\% (Figure~\ref{fig:results_scenes_mobile}). We are again able to keep power consumption within our budget with image quality very close to the maximum quality.

\section{Discussion}
\label{sc:discussion}
Our on-the-fly power-aware rendering framework successfully addresses the two key limitations of previous work: it does not require any precomputation, and it can handle dynamic scenes. We have shown results for four different scenes of different characteristics, demonstrating large power savings while maintaining image quality close to maximum quality.
Analysing the optimal configurations chosen by our framework, we notice that image resolution is rarely lowered, since it leads to high quality errors. Our algorithm does not lead to any degradation of the frame rate, as we demonstrate in our supplemental video.


Our power prediction model may have other applications beyond budget rendering. For example, by detecting an increase in power consumption (which indicates higher rendering complexity), it could analyse different options to avoid frame-rate drops in video games before they happen. It could also be applied to estimate the total energy consumption of a new rendering task, given a limited set of initial data. Since $b$, $t$, and $f$ in Equation~\ref{eq:power_prediction_model_extended} can be fetched with native OpenGL queries, incorporating our power prediction model to any project is straightforward, and requires no modifications of any existing shader code.

\begin{figure}
    \includegraphics[width=\columnwidth]{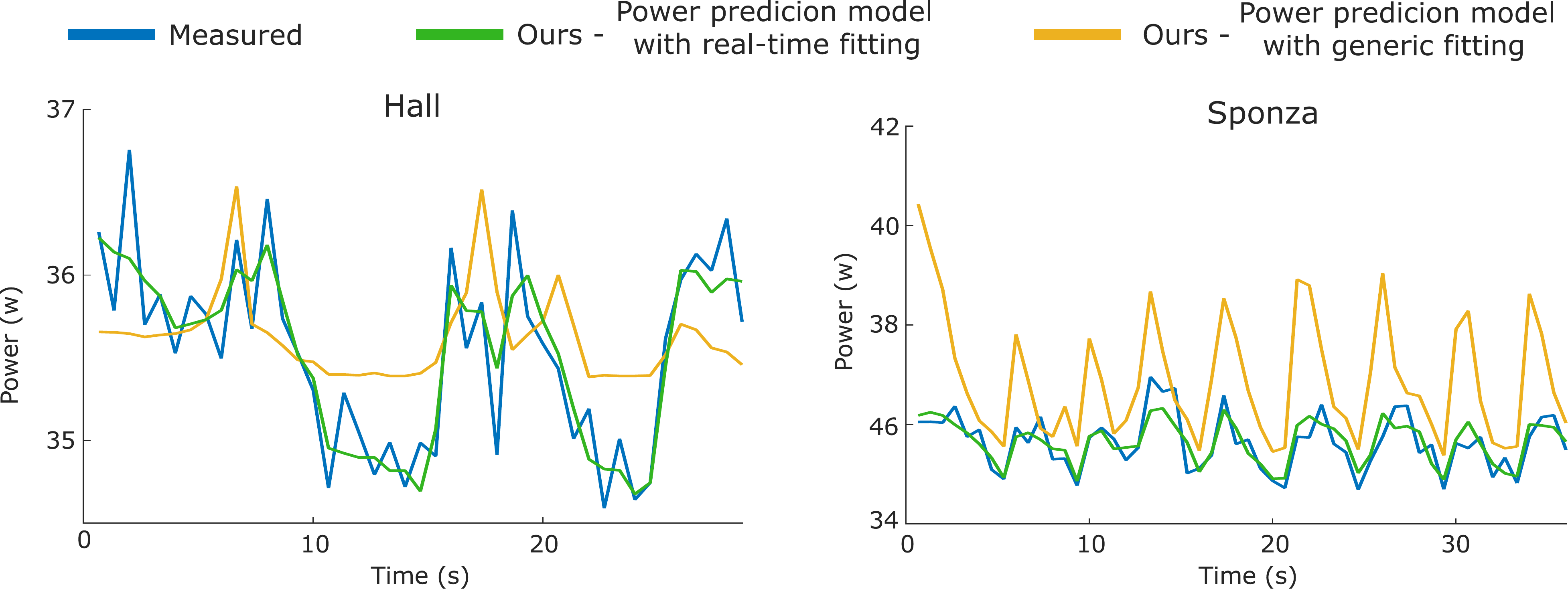}\label{fig:baseline_fitting}
    \caption{\new{Power cosumption of the Hall and Sponza scenes with ground truth measured data, predicted with our power model with real-time fitting and  predicted with our power model with generic fitting.}}
\label{fig:power_model_discussion}
\end{figure}

As we have shown, real-time fitting of our power prediction model provides very high accuracy. However, it is also possible to fit the model with a generic dataset to obtain valid coefficients before running any specific scene. To do that, we collect rendering samples from dummy scenes with a varying number of batches, vertices, and fragments, covering the whole parameter space from $P_m$ to $P_M$. With these data, we fit $k_{bi}$, $k_{vi}$, and $k_{fi}$ in Equation~\ref{eq:power_prediction_model_extended} using linear regression on the power consumption and number of primitives. This offline process takes around 4 minutes for one rendering configuration, and provides a reasonable approximation of the actual power consumption (see yellow curve in Figure~\ref{fig:power_model_discussion}). 

To perform our runtime quality error computations, we have set a frequency of 10 frames, which we have selected to be as small as possible without minimizing the impact of background rendering on the frame rate. Alternatively, this frequency could be automatically adjusted during runtime according to the current frame, to guarantee a given target frame rate.

Throughout the whole paper, we have defined the power optimal configuration as the one with lowest quality error that meets our power budget (Equation~\ref{eq:problem}). Similar to Wang's work~\shortcite{Wang2016}, solving the analogous problem of obtaining the rendering configuration with the lowest error consumption within an error budget is straightforward:
\begin{align}\label{eq:problem_error}
\mathbf{s} & = \argmin_\mathbf{s} P(\mathbf{s},\mathbf{c}) && \text{subject to} && e(\mathbf{s},\mathbf{c}) < e_{\text{bgt}}
\end{align}
In this case, we would start our selection of the optimal configuration by estimating the error for all configurations and discarding the ones above the budget, then predicting power for the remaining configurations, and choosing the one with lowest consumption. 

\textbf{Limitations and future work:} Our framework still has some limitations that could be addressed in future work. Our power prediction model seamlessly supports dynamic scenes, by using the information of the current frame to predict power consumption. However, our error computation mechanism needs to explicitly store the positions of moving objects and restore them for background rendering. For scenes with a large number of moving objects, this could become too computationally expensive. 

\new{Our power model is based on the typical rendering pipeline with basic processing of batches, vertices, and fragments. It does not accurately model other GPU stages that could be integrated into the pipeline, such as geometry shaders or tesselation, which should be included as additional contributors to our formula. Apart from that, our model is already able to seamlessly represent the additional fragments generated by a geometry shader.}

\newreb{
We have demonstrated the viability of our framework using a reasonable number of different shaders, under the strict constraint of real-time execution. We have not, however, exhausted all the possibilities; testing our proposal in a complex rendering engine is a very interesting direction for future work.
}

Our framework may produce inaccurate predictions when the rendering samples used to fit the model do not include information related to a certain pass (e.g., the \textit{Reflections} pass if no reflective surfaces were being rendered at the time). However, these inaccuracies tend to last only a few frames, and the system eventually self-corrects; we have found that this does not have a relevant impact on performance in the long run.

\section*{Acknowledgements}

We would like to thank all reviewers for their insightful comments. We also thank Bowen Yu for his contribution in the initial phase of this project, and Julio Marco for helping with figures and proofreading the paper. This research has been partially funded by National Key R\&D Program of China (No. 2017YFB1002605), NSFC (No. 61472350), Zhejiang Provincial NSFC (No. LR18F020002), the Fundamental Research Funds for the Central Universities (No. 2017FZA5012), European Research Council (ERC) under the European Union's Horizon 2020 research and innovation programme (CHAMELEON project, grant agreement No 682080), and the Spanish Ministerio de Econom\'ia y Competitividad (projects TIN2016-78753-P and TIN2016-79710-P).

\bibliographystyle{eg-alpha-doi}
\bibliography{references}

\newpage

\end{document}